\pdfoutput=1
\documentclass[format=acmsmall, review=false, authorversion, screen=true]{acmart}

\usepackage{booktabs} 
\usepackage{fancybox}
\usepackage{enumitem}

\setlist[itemize,1]{label=$\blacktriangleright$}

\usepackage[ruled]{algorithm2e} 

\SetAlFnt{\small}
\SetAlCapFnt{\small}
\SetAlCapNameFnt{\small}
\SetAlCapHSkip{0pt}
\IncMargin{-\parindent}

\begin{document}

\title[The Trusted Edge]{The Trusted Edge}

\subtitle{\textit{``From non-starter to contender: making Edge Computing more appealing for business through intellectual property protection in open ecosystems.''}} 

\author{Christian Meurisch}
\affiliation{%
  \institution{Technical University of Darmstadt}
  \streetaddress{Hochschulstra{\ss}e~10}
  \city{D-64289 Darmstadt}
  \country{Germany}
}
\email{meurisch@tk.tu-darmstadt.de}

\maketitle

\renewcommand{\shortauthors}{C. Meurisch}

Edge computing promises to reshape the centralized nature of today's cloud-based applications by bringing computing resources, at least in part, closer to the user~\cite{shi2016promise}. 
Reasons include the increasing need for real-time (short-delay, reliably-connected) computing and resource-demanding artificial intelligence (AI) algorithms that overstrain mobile devices' batteries or compute power but are too bandwidth-demanding to be offloaded to a distant cloud~\cite{zhou2019edge}. 
For instance, autonomous driving/flying, 3D~urban modeling, and augmented/virtual reality (AR/VR) applications are newly emerging use cases that would greatly benefit or be first made possible by edge computing~\cite{qiu2018avr,wang2018service}.

Although the roots of edge computing go back to the late 1990s (with a conceptual foundation based on `cloudlets' in 2009~\cite{satyanarayanan2017emergence}) and large consortia/initiatives like OpenFog and MEC (mobile edge computing) arose lately to push standardization forward~\cite{wang2017survey}, edge computing is not yet largely integrated into business models (if any, only in small company-owned ecosystems). 
Besides the complexity of management, which receives a lot of attention in research~\cite{roman2018mobile}, two other `showstoppers' indicated in~\cite{satyanarayanan2017emergence} by Satyanarayanan block the breakthrough of edge computing: 
(1)~a bootstrapping problem---without large-enough deployments of available edge resources, there is little incentive for companies to create such decentralized new applications, and vice versa; 
(2)~the weaker security measures than `the cloud'---mainly related to tampering and spying~\cite{roman2018mobile}.

As the past has shown (e.g.,~the rise of the Internet), the most promising way to tackle Issue\,1 is to foster an open edge computing ecosystem, which would attract investments from multiple parties while minimizing the individual risks. 
On the downside, companies may need to run their business logic on (untrusted) third-party devices (e.g.,~in urban infrastructures~\cite{muehlhaeuser2021street,baumgartner2017emergency,nguyen2017adaptive}), leading to aggravated security issues\footnote{https://www.kaspersky.com/blog/secure-futures-magazine/edge-computing-cybersecurity/31935/} (see Issue\,2): 
the risk of leaking confidential data, such as protected AI algorithms/models representing the companies' intellectual property (IP)~\cite{ma2020security}, is incalculable from today's perspective, making edge computing a non-starter for most business applications.
A recent study found that two-thirds of IT teams saw edge computing as more of a threat than an opportunity;
more than half ($52\%$) see the threats as rooted in security concerns\footnote{https://www.techrepublic.com/article/66-of-it-teams-view-edge-computing-as-a-threat-to-organizations/}.
To me, the serious problem of IP protection in particular is one of the major challenges, which is largely underestimated, even ignored, especially in edge computing research.

\section*{The Need for Trusted Edge Computing (TEC)}
One promising research direction to alleviate this problem is based on \textit{trusted computing}~\cite{mitchell2005trusted}, whereby the code is attested by a (hardware-based) trusted platform module to detect unauthorized changes and ensure consistent behavior in expected ways. 
Remote attestation (RA) is one path to securely verify the \textit{integrity} of the code on an untrusted device (\textit{prover}) by an external/distinct party (\textit{verifier}).
RA can potentially be performed over an open multi-hop network with public-key encryption required for \textit{authenticity}~\cite{carpent2017lightweight}. 
Since the initial content is usually loaded from unprotected memory of the device, code integrity alone is not sufficient to keep the companies' protected business logic confidential. 
First approaches (e.g.,~\cite{brasser2018voiceguard}) exploit the secure channel established for RA to subsequently load confidential parts of the companies' applications directly into the protected memory of the device~\cite{henson2014memory}, ensuring \textit{confidentiality}.

Although those `traditional' hardware-based attestation concepts may be a good starting point and became widely available for different processor architectures (e.g.,~Intel SGX\footnote{https://www.intel.com/content/www/us/en/architecture-and-technology/software-guard-extensions.html}, ARM TrustZone\footnote{https://www.arm.com/why-arm/technologies/trustzone-for-cortex-a/tee-and-smc}) or cloud environments (e.g.~Microsoft Azure\footnote{\label{myfootnote}https://azure.microsoft.com/overview/azure-ip-advantage, https://azure.microsoft.com/solutions/confidential-compute}), they are not directly applicable to edge computing with its inherent decentralized nature, heterogeneous devices, and real-time demands~\cite{abera2016things}. 
Close research cooperation between both communities, the `edge computing' community and the `security \& privacy' (or more specifically the `trusted computing') community is therefore indispensable. 
Let us coin this interdisciplinary cooperation as \textit{trusted edge computing} (TEC) with a prioritized overarching goal of developing concepts and methods to protect the business logic of application providers (and thereby their IP) that are specially tailored for open edge computing ecosystems. 

\section*{Open Challenges}
Accordingly, TEC is confronted with several inherent challenges including \textit{hardware} and \textit{connectivity heterogeneity}, \textit{support for specific platform concepts and services}, \textit{performance}, and \textit{scalability}.

\subsection*{Hardware Heterogeneity}
In edge computing, heterogeneous environments consisting of devices with different hardware and communication specifications are quite common. Especially for resource-constrained devices with an often missing trust anchor due to cost and complexity, lightweight and thus more challenging RA techniques need to be applied and further developed: \textit{software-based RA}---only providing security guarantees when strong but partly unrealistic assumptions about adversarial behavior are made---or \textit{hybrid-based RA}---representing a hardware/software co-design that requires at least some hardware features (e.g.,~memory isolation, exclusive secret key access)~\cite{carpent2017lightweight}. 

\subsection*{Connectivity Heterogeneity}
Assuming a 3n-tier edge computing architecture\footnote{The newly emerging 3n-tier architecture extends the traditional ``device-cloud'' paradigm to a three-tier ``device-edge-cloud'' setting (vertical), where additionally n interconnected edge nodes participate in the middle tier (horizontal).}, participating edge devices may only have a multi-hop connection to the verifier, which is not necessarily reliable. 
Since software-based RA is not viable in such network settings due to the variable round-trip delays and the resulting skewed time measurements~\cite{abera2016things}, hybrid-based RA is minimally necessary to ensure confidentiality; 
i.e.,~resource-constrained edge devices should, therefore, have at least a minimum set of aforementioned hardware features. 
Whether hybrid- or hardware-based RA, in both cases appropriate defense mechanisms against possible physical adversaries, decentralized network and side-channel attacks must be incorporated, either by adapting existing approaches (e.g.,~\cite{oleksenko2018varys}) or by developing specially designed approaches for edge computing environments.

\subsection*{Support for Specific Platform Concepts and Services}
TEC must also support specific platform concepts for edge computing and services.
Services can be developed as a set of small individual functions, modules or even other services~\cite{meurisch2019assistantgraph} that are instantiated (in ms) and executed on demand~\cite{boucher2018putting}. 
These raise several challenges for TEC including (1)~the need for runtime attestation to dynamic (un-)load modules at runtime~\cite{abera2016things}, and (2)~minimizing the size of the trusted computing base e.g. inside an SGX enclave due to the limited protected memory between 64-128\,MB (depending on the hardware and implementation) and the resulting expensive swapping. 
For instance, container-based virtualization therefore requires lightweight TEC mechanisms~\cite{arnautov2016scone}, e.g.,~only loading the confidential service parts of a container into the enclave.

\subsection*{Performance Issues}
TEC mechanisms, in turn, will inevitably lead to different performance overheads: 
(i)~computational overheads both during the initialization process and at runtime, especially on resource-restricted edge devices~\cite{kaup2018progress}; 
(ii)~overhead caused by employing protection mechanisms against side-channel attacks~\cite{brasser2018voiceguard}. 
Due to the strict responsiveness requirements of edge computing, it is critical to achieve acceptable overheads and improve the tradeoff between security and performance.
For instance, the dynamic provisioning of services or modules requires fast setup times in the range of micro- to milliseconds~\cite{boucher2018putting}; 
this can be realized by, among others, various pre-initialization stages like preloading encrypted AI algorithms/models~\cite{meurisch2020privacy} or partitioning neural networks computation across multiple enclaves~\cite{elgamal2020serdab}.
Another way to improve responsiveness is the use of hardware accelerators such as GPUs and FPGAs for resource-intensive services, e.g.,~relying on deep learning models.
These hardware accelerators become more and more an essential capability in edge computing, but their trust model is typically crude in a multi-tenant environment.
Supporting trusted execution environments on GPUs of edge devices is an open challenge that TEC must address.~\cite{volos2018graviton}

\subsection*{Scalability Issues}
Current RA techniques are often limited to a single-prover/-verifier scenario, which would not scale well in decentralized edge computing setups with many devices (referred to as \textit{device swarms}). 
In particular, open specific challenges lie in designing RA mechanisms that must (i)~be efficiently scalable to handle multiple edge devices in a typical 3-tier (or emerging 3n-tier) architecture, possibly with the support of a centralized trusted resource management as an `anchor', e.g., by using Microsoft Azure\textsuperscript{\ref{myfootnote}}, (ii)~cope with the potentially dynamic topology due to user mobility, and (iii)~be independent of the underlying integrity measurement mechanisms used by the heterogeneous devices~\cite{abera2016things}. 
To advance such swarm attestation, first approaches (e.g.,~\cite{carpent2017lightweight}) rely on purpose-built hash functions and/or (recursive) measurement aggregations. 
Further considerations such as caching of the encrypted functions at the edge and their distribution ways must be included in RA techniques to specifically fit into existing edge computing provisioning and resource management methods.

\section*{Looking Ahead}
In addressing these challenges, TEC would not only enable IP protection for the providers but also privacy-preserving processing of (sensitive) end-user data in the next breath~\cite{meurisch2019assistantgraph,meurisch2017extensible}. 
For instance, TEC solutions may complement or make emerging `bring your own data' (BYOD) concepts, such as the Databox~\cite{perera2017valorising} and PDSProxy~\cite{meurisch2020pdsproxy} concepts, practically applicable at all: they can allow a user to personalize untrusted (third-party) devices and their respective applications in a confidential and ad-hoc manner -- a comprehensive literature survey is provided by~\cite{meurisch2021data}.

\subsection*{TEC may break down the entry barrier for business} 
This article makes the case for urgently needed interdisciplinary research in the sub-discipline of trusted edge computing, which goes far beyond authenticity and integrity through attestation: 
it particularly needs to focus on (1)~\textit{confidentiality} to protect the IP of companies and (2)~the \textit{decentralized nature of heterogeneous edge nodes} to efficiently establish trust in those remote (third-party) devices. 
Otherwise, edge computing risks being a non-starter for business: current research priorities are heading towards a situation of sophisticated, scalable computing platform concepts, but probably companies hesitate to use them -- despite the demand for new applications -- due to the inadequate and neglected protection of intellectual property, especially in open infrastructures.

\clearpage
\bibliographystyle{ACM-Reference-Format}
\bibliography{trustededge}

\end{document}